# Ferroelectric Quantum Point Contact in Twisted Transition Metal Dichalcogenides

Wei Ren[1], Shiyu Guo[1], Daochen Long[1], Noah Friedman[1], Jingyuan Xian[1], David Li[1], Jack Tavakley[1], Jeongsoo Park[1], Kenji Watanabe[2], Takashi Taniguchi[3], Ke Wang[1*]

[1]*School of Physics and Astronomy, University of Minnesota, Minneapolis, MN, USA*

[2]*Research Center for Electronic and Optical Materials, National Institute for Materials Science, Tsukuba, Ibaraki, Japan*

[3]*International Center for Materials Nanoarchitectonics, National Institute for Materials Science, Tsukuba, Ibaraki, Japan*

**In twisted transition metal dichalcogenides (tTMDs), atomic reconstruction gives rise to moiré domains with alternating ferroelectric polarization, whose domain size and overall electric dipole moment are tunable by an out-of-plane electric field. Previous transport measurements in Hall bar devices have successfully demonstrated the overall ferroelectric behavior of tTMDs from a collective ensemble of ferroelectric moiré domains. To locally probe a single ferroelectric moiré domain, we fabricate and study mesoscopic quantum transport via a gate-defined twisted molybdenum disulfide (tMoS$_2$) quantum point contact (QPC). The local property of a single moiré domain is invulnerable to long-range disorder and twist-angle inhomogeneity, resulting in an unusually long conductance plateau with large electrical hysteresis. The comparison between local and global measurements confirms that antiferroelectricity can emerge from alternating polarization of individual ferroelectric domains. Using a QPC as a single charge sensor, we characterize the nature and time scale of different domain evolution mechanisms with single atomic dipole resolution. Our findings shed new light on the microscopic ferroelectric behavior and dynamics within a single tTMD moiré domain, paving the way toward more advanced ferroelectric quantum devices with tunable local Hamiltonian, such as ferroelectric tTMD quantum dots (QDs).**

In many van de Waals (vdW) materials, the lack of inversion symmetry can give rise to spontaneous electric polarization and ferroelectricity [1,2]. For non-ferroelectric 2D materials, such inversion-symmetry breaking can be engineered by stacking with a twist [3–9]. This approach is also widely used to facilitate exotic quantum phenomena including unconventional superconductivity [10–16], correlated insulators [17–22], quantum anomalous Hall effects [23–

*Corresponding author: kewang@umn.edu

25], etc. For small twist angles, the twisted-bilayer undergoes substantial atomic reconstruction [26–29], resulting in moiré domains with alternating atomic (AB and BA) stacking orders and opposite spontaneous electric polarizations. Under external out-of-plane electrical field $E_0$, the moiré domains of aligning polarization expand while the neighboring domains with opposite polarization shrink in size (known as moiré ferroelectricity [1–4,7] or moiré domain antiferroelectricity [30]), and the material behavior as a function of $E_0$ becomes hysteretic. Such hysteresis has been demonstrated by several pioneering transport measurements of tTMD Hall bar device geometries [7,31], successfully confirming the overall ferroelectric behavior. However, in the presence of twist angle inhomogeneity and atomic disorders, the resulting overall ferroelectric transport behavior is a collective average over alternating domains of varying size and local properties. Electrical characterization and control of local ferroelectric behavior at the single moiré domain scale can shed new insights on the local microscopic mechanism of moiré ferroelectricity, and provide technical components for next-generation quantum devices with versatile electrical manipulation over novel quantum degrees of freedom [32–34].

To measure the electric properties of a single ferroelectric moiré domain and to investigate the local microscopic mechanism of moiré ferroelectricity, we fabricate gate-defined QPCs in twisted $MoS_2$ and conduct low-temperature mesoscopic quantum transport measurement. Figure 1a shows the schematic of the device architecture. A piece of hexagonal boron-nitride (hBN), a pair of pre-doped few-layer graphite contacts (FLGs) [35], two pieces of few-layer $MoS_2$ with a small twist angle <1°, and another piece of hBN are picked up in series. The vdW heterostructure is then transferred onto a series of atomically clean local bottom gates (see Methods) pre-deposited on a $SiO_2$/Si substrate (figure 1b). The device is subsequently dry-etched to remove material outside the local-gated region, and to ensure the mesoscopic transport (inset, white arrows) via the gate-defined QPC as the sole conducting channel.

**Ferroelectric Quantum Point Contact**

We measure the electrical transport signal of the tMoS$_2$ QPC device at 4 K (figure 2a). Equal gate voltages ($V_{QPC}$) are applied to a selected pair of the local bottom gates, to deplete the electrons in region directly above the gates, and to electrostatically define a 1D confinement (QPC) in regions above its gate separation. The width of the 1D conducting channel is adiabatically tuned from ~100 nm all the way to zero (pinch-off) with $V_{QPC}$ through the local fringing fields and become comparable to moiré lattice constant near the pinch off. Therefore, mesoscopic transport characterizes the local properties of one or a few moiré domains. The silicon back gate voltage ($V_g$) simultaneously tunes two experimental parameters in the QPC channel: (1) the out-of-plane electric field $E_0$ that tunes the electric polarization and (2) the carrier density that determines the Fermi wavelength. The device lacks a global top gate (and is not dual-gated), as the top-gate voltage will dope the region above the QPC gates which should stay depleted and lift the 1D constriction necessary for local mesoscopic transport. The capability of individually tuning the carrier density and electric field in the channel requires local top and bottom nano-gates that are perfectly overlay-aligned, which is beyond the scope of this work. The regions directly above the metal QPC gates can deplete charge carriers for $V_{QPC} < -5$ V and therefore transport across the device is forced to go through a mesoscopic constriction (QPC) at the center of the device (figure 2a, blue). To ensure such 1D confinement exists independent of the applied $V_g$, we employ the pre-deposited local QPC back gate (instead of the post-deposited local QPC top gate) that screen $E_0$ for the depleted region.

Figure 2b shows the measured four-probe resistance of the device as a function of $V_g$ at fixed local gate voltage of $V_{QPC} = -7$ V. A large electric hysteresis of $\Delta V_g \sim 120$ V has been observed between forward and backward scans of $V_g$ (figure 2b, with arrows indicating the scan directions), independent of the back gate scan rate $dV_g/dt$. This confirms its origin as ferroelectricity, instead of capacitive circuit components or charge traps, typically associated with a RC time-constant that sensitively depends on scan rates [36–38] (also see SI section S2). Two to three conductance plateaus are visible in both the forward and backward scans, though its conductance is not quantized as the bulk two-dimensional electron gas (2DEG) resistance leading to the QPC is comparable to the ballistic QPC resistance, resulting from significantly lower carrier mobility (compared to graphene, GaAs and Si 2DEGs). Similar observations have been previously reported in TMD QPCs [35], and the conductivity of 2DEG in tMoS$_2$ moiré superlattice is expected to be further limited by the additional twist-angle inhomogeneity.

The width of each identifiable conductance plateau is similar to previously reported TMD QPCs [35,39,40], except for the unusually-wide first plateau observed in forward scan. We show that this is consistent with the physic picture of electrons in a QPC being simultaneous subject to two types of electric fields, (1) the external electric field $E_0$ (from back gate), which has one-on-one linear relationship with $V_g$, and (2) an effective molecular field $E'$ proportional to the electric polarization $P$, which is hysteretic due to the ferroelectricity of moiré domains near the QPC.

Since carriers in the QPC are simultaneously tuned by $E_0$ and $E'$ and the former is directly given by non-hysteretic linear relation $E_0 = \frac{c_g}{\varepsilon_{hBN}\varepsilon_0}V_g$ (see SI section S3), the hysteretic function of $E'(V_g)$ is responsible for the observed hysteresis of $\sigma(V_g)$ (figure 2b). Here, $c_g$ is the capacitance per unit area between the silicon back gate and the tMoS$_2$ 2DEG, $\varepsilon_{hBN}$ is the relative permittivity of hBN, and $\varepsilon_0$ is the vacuum permittivity. And because $E'$ and $E_0$ are proportional to $P$ and $V_g$, respectively, $E'(V_g)$ can be understood with textbook hysteresis loop of $P(E_0)$, as illustrated in figure 2c. The blue arrows depict the direction of dipoles at saturation and at zero $P$ or $E_0$, while the colored points in figure 2c shows the instantaneous electrical polarization at the matching points of QPC conductance measurement in figure 2b.

Starting with a large positive $V_g$ (blue diamond) with a large upward external electric field $\vec{E}_0$ ($E_0 > 0$ is defined as pointing upward), where all dipoles (blue arrows in figure 2c) are polarized upwards along the external field direction (near saturation), creates an effective internal molecular field $\vec{E}'$ in the downward direction, and the net electric field experienced by the QPC channel, $\vec{E}_{net}$ is given by: $\vec{E}_{net} = \vec{E}_0 + \vec{E}'$ ($E_{net} = E_0 - E'$). For the ease of discussion, we define parameter $\beta$ as the partial derivative of the molecular field with respect to the external field: $\beta = \frac{\partial E'}{\partial E_0}$, whose value is proportional to the local slope $\frac{\partial P}{\partial E_0}$, in the $P(E_0)$ hysteresis loop. The carrier density in the QPC channel can be qualitatively expressed by $n = CE_{net} = C(E_0 - E')$, where $C$ is a constant. The rate at which the carrier density is tuned by $E_0$ is therefore given by its derivative, $\frac{\partial n}{\partial E_0} = C(1 - \beta)$.

As we reduce the size of positive $E_0$ with a scan of $V_g$ from the blue diamond to the red circle, $P$ has a very weak dependence on $E_0$ and $V_g$, and therefore $\beta \sim 0$ and $\frac{\partial n}{\partial E_0} \propto C$. The carrier density in QPC channel is tuned effectively as if the ferroelectricity is absent, and it rapidly pinches off as a function of $V_g$ after exhibiting a few conductance plateaus of standard sizes. As we continue the backward scan of $V_g$ from the red circle all the way to a large negative value (yellow triangle), the process corresponds to traversing the sharp downward slope of the $P(E_0)$ hysteresis loop, with rapidly changing polarization and $\beta \sim 1$. As a result, the further depletion of charge carrier density

(Fermi level further into the bandgap) of the QPC channel is not effectively tuned by $V_g$, with a reduced rate of $\frac{\partial n}{\partial E_0} = C(1-\beta) \sim 0$, just a small fractional of $C$. As a result, despite the very negative $V_g$ value at the yellow triangle, the QPC is not deep into the pinch off, and the Fermi level is just slightly lower than the conduction band minima compared to red circle.

Now we start the forward scan of $V_g$, from the yellow triangle to cyan square, traversing through a flat part of $P(E_0)$ hysteresis loop. The weak dependence of $P$ on $E_0$ once again gives $\frac{\partial n}{\partial E_0} \sim C$, and the QPC channel quickly becomes conducting due to (1) the uncompromised capacitive coupling to $V_g$, and (2) the QPC being near pinch off at yellow triangle to begin with. Traversing the sharp slope of $\beta \sim 1$ on the $P(E_0)$ curve, from the cyan square to the purple star, the effectiveness of tuning carrier density in the QPC channel with $V_g$ is once again significantly reduced to $\frac{\partial n}{\partial E_0} = C(1-\beta) \sim 0$, resulting in the QPC remaining in the first conductance plateau for an usually large range of $V_g$. As $\beta$ reduces from ~1 to ~0, from purple star to blue diamond, the QPC eventually gets tuned out of the first plateau, and the subsequent higher plateaus resume normal widths due to regained capacitive effectiveness of $V_g$.

Figure 2d shows the measured conductance during forward and backward voltage sweeps at different symmetric scan ranges. Larger scan ranges produce increasingly pronounced hysteresis loops and extended conductance plateaus. The systematic enlargement of hysteresis loops with increasing scan range suggests the presence of unsaturated ferroelectric polarization within the experimentally accessible electric field range ($V_g$ from -80 V to +80 V).

**Probing Domain Evolution with Single Dipole Sensitivity via Gate-defined QPC**

At the steep slope between conductance plateaus (or pinch-off), the QPC conductance is extremely sensitive to local electrostatic environment, capable of detecting the slight change in Coulomb potential from a single electron transport or displacement nearby. The QPC as a single charge sensor has been widely used in state-of-the-art Loss-DiVincenzo qubits [41] for high fidelity single-shot qubit read-out [42–45]. In this work, we leverage the electrostatic sensitivity of gate-defined tMoS$_2$ QPC to probe the evolution of the ferroelectric moiré domain with single dipole sensitivity. From pinch off (figure 2b, yellow triangle), we ramp $V_g$ forward adiabatically until the QPC conductance reaches halfway to the first conductance plateau (figure 2b, cyan square).

After the QPC reaches half the conductance of the first plateau, we keep $V_g$ constant to park the QPC at the steep slope of its conductance $\sigma(V_g)$ and measure the four-probe resistance of the device as a function of time (figure 3a). As the size of the dominating (dominated) domain with downwards (upwards) electric polarization gradually decreases (increases) its domain size towards equilibrium over a course of few hours, the total magnitude of $P$ and $E'$ decreases, $E_{net}$ becomes more negative, and the QPC conductance decreases as it is tuned closer to pinch-off. An exponential dependence is expected from an assembly of non-interacting dipoles evolving to an equilibrium configuration, while a stretched exponential response, or a quasi-logarithmic dependence, is previously reported in ferroelectric thin films exhibiting slow relaxation phenomena, indicating the presence of strong dipole-dipole interaction [46,47]. Qualitatively, time dependence observed here is dominated by exponential behavior, with only a slight slow-down from possible logarithmic components (See figure S3, where the same time-dependence is replotted on a semi-log scale). This suggests that the electric dipoles in tMoS$_2$ are only weakly interacting and largely independent. This confirms that moiré ferroelectricity arises from atomic reconstruction of moiré superlattice instead of its electronic reconstruction, and its microscopic mechanism spontaneous inversion-symmetry breaking of relaxed moiré domains instead of flat-band induced electron dipole-dipole correlation. The time-dependence is fitted with $\sigma = \sigma_0 e^{-t/\tau}$, with extracted characteristic time scale of $\tau = 1.18$ h, which have been qualitatively reproduced in multiple devices (see SI section S7).

Figure 3b shows a zoom scan during the first 0.3 h of figure 3a. The QPC conductance is observed to exhibit telegraph switching behavior between two floor values (marked by green dashed curves), each demonstrating the overall near-exponential dependence with same characteristic $\tau = 1.18$ h. The conductance difference between the two floor values is consistently around 0.02 uS throughout the scan range and one order of magnitude larger than the characterized noise level of 0.002 uS. This is consistent with events of single-dipole switching at the domain boundary located in close proximity to the QPC, demonstrating the single-dipole sensitivity of tMoS$_2$ QPC. The average characteristic timescale observed for single-dipole flipping is on the order of 1 min.

**Overall tMoS$_2$ Material Property**

To investigate the global transport behavior of the entire tMoS$_2$ sample, we simultaneously tune the global back gate ($V_g$) and the local bottom gate ($V_{QPC}$) voltages, with step sizes proportionally to their capacitive coupling of $\Delta V_g/\Delta V_{QPC} = c_{hBN}/c_g$ (see SI section S3). Here, $c_{hBN}$ is the capacitance per area between the local gates and 2DEG. This ensures the carrier density across the entire 2DEG is approximately the same, and electrical transport occurs across the entire sample in the absence of gate-defined constriction. Figures 4a, b show the measured four-probe resistance of two different devices (tMoS$_2$ sample #1 and sample #2) as a function of $V_g$ for different forward and backward scan ranges. Compared to the mesoscopic measurement via an individual (or few) moiré domain near the QPC (figure 2), a qualitative difference is observed in the collective transport behavior of entire device sampling all its moiré domains. First, the size of the hysteresis loop is significantly smaller when averaging over a large ensemble of domains (up to ~60 V with in the experimental accessible range) compared to an individual domain (up to ~100 V), as expected from realistic twist-angle inhomogeneity and confirming the advantage of mesoscopic measurement in probing pristine moiré ferroelectricity. This observation has been reproduced in multiple control devices (see SI section S7). Second, the observed hysteresis loops are now located entirely in the $V_g < 0$ or $V_g > 0$ regions, instead of spanning across $V_g = 0$.

In a conventional ferroelectric material, system dimension $L \gg a$, in which $a$ is atomic length scale between nearest-neighbor anti-aligned electric polarizations. Similarly, at the limit where $L$ (device dimension) $\gg L_{moiré}$ (moiré lattice constant) and $L_{moiré}$ being constant, the alternating ferroelectric domain with opposite electric polarization can give rise to anti-ferroelectricity of the overall polarization as a function of $E_0$ (figure 4c), with signatures of double hysteresis loop in global transport behavior (figure 4a). Two anti-ferroelectric hysteresis loops (figure 4a) can be identified in region $E_0 < 0$ (or $E_0 > 0$) away from (without crossing over) $E_0 = 0$, in contrast to the transport signature of a single ferroelectric domain which spans across and roughly centered around $E_0 = 0$ (figure 2). The global anti-ferroelectric hysteresis loops located within the $E_0 < 0$ region are observed for multiple devices measured (figure 4a, 4b), and are significantly smaller in size and shifted away from $E_0 = 0$ compared to the transport signature of a single ferroelectric domain, as expected from the moiré domain anti-ferroelectricity. However, the anti-ferroelectric hysteresis loop located at $E_0 > 0$ shows significant device to device variation, that can be attributed to realistic twist angle inhomogeneity which results in spatially varying moiré domain sizes. At low carrier density ($E_0 < 0$), a single electron dipole exists within each domain independent of domain size, the alternating electric polarizations are equal in size, analogous to a conventional anti-ferroelectric material. At higher carrier density ($E_0 > 0$), the variation in moiré domain sizes results in variation of number of charges carriers in each domain and the resulting electric dipole moment, the alternating electric polarizations are no longer equal in size and overall electric polarization are determined by that of the larger domains, exhibiting ferroelectric hysteresis loop with its signature plateaus. Future experiments combining the direct imaging of moiré domains and local and global transport measurements may shed new light in understanding how the overall ferroelectric properties of tMoS$_2$ quantitatively depend on spatial-homogeneity and the size of moiré lattice constant. Figure 4d shows the time evolution of the global conductance of the device, parked at the sharp slope of the hysteresis loop. The absence of single dipole switching behavior (figure 4e) confirms that all moiré domains contribute to the transport signature of the global measurement (figure 4a), and that the sharp slope in the conductance belongs to the anti-ferroelectric hysteresis loop (instead of conductance of an accidentally-formed QPC), making the global measurement insensitive to a single dipole switching event.

**Conclusion**

In summary, we investigate mesoscopic quantum transport via gate-defined tMoS$_2$ quantum point contacts and demonstrate unusually long and hysteretic conductance plateau from ferroelectric response of a single moiré domain. By operating the QPC as a single-dipole sensor, we characterize the time scale of the ferroelectric domain evolution and single-dipole switching events. In comparison, we observe distinctly different transport signatures in macroscopic transport without the gate-defined 1D QPC confinements, consistent of the overall anti-ferroelectricity from a large number of ferroelectric moiré domains of alternating polarizations. Our works provide a new approach in measuring the transport properties of a single moiré domain with single-dipole sensitivity, invulnerable to long-range disorder and twist-angle inhomogeneity, shedding new light on microscopic transport physics of a single moiré unit cell that can't be accessed with a conventional Hall bar device. For future experiments, more advanced device architecture with perfectly aligned local bottom and top gates can allow independent control over local carrier density and external electric fields. Direct imaging techniques such as near-field scanning optical microscopy (NSOM) or scanning tunneling microscopy (STM) can also be used to confirm the length scale and spatial homogeneity of individual ferroelectric domains, and its implications to the ferroelectric mesoscopic transport and anti-ferroelectric macroscopic transport signatures.

**Acknowledgements**

We thank Yi Huang, Jihang Zhu, Alex Kamenev and Dan Dahlberg for useful discussions. The work at UMN was supported by the National Science Foundation CAREER Award NSF-1944498. Portions of the UMN work were conducted in the Minnesota Nano Center, which is supported by the National Science Foundation through the National Nano Coordinated Infrastructure Network (NNCI) under Award Number ECCS-1542202. Portions of the hexagonal boron nitride material used in this work were provided by K.W and T. T. K.W. and T.T. acknowledge support from the Elemental Strategy Initiative conducted by the MEXT, Japan (Grant Number JPMXP0112101001) and JSPS KAKENHI (Grant Numbers 19H05790, 20H00354 and 21H05233).

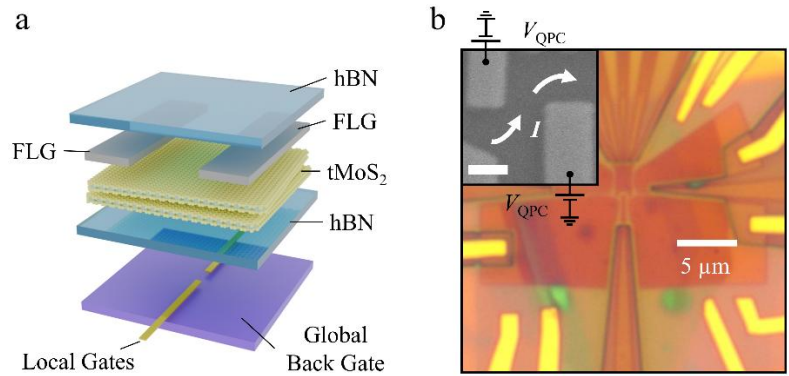

**Figure 1 | Gate-defined Twisted MoS$_2$ Quantum Point Contact.** (a) Schematic device architecture. Twisted molybdenum disulfides (tMoS$_2$) are contacted by few-layer graphite flakes (FLGs), which are then encapsulated by hBN and transferred onto local bottom gates pre-deposited on Si/SiO$_2$ substrates. (b) Optical microscope image of a typical device with multiple quantum point contacts (QPCs) and scanning electron microscope (SEM) image (inset) of a typical pair of the local gates, electrostatically forming an electrostatically defined 1D constriction (QPC) for the current (white arrow) to flow through. The scale bar in the inset is 200 nm.

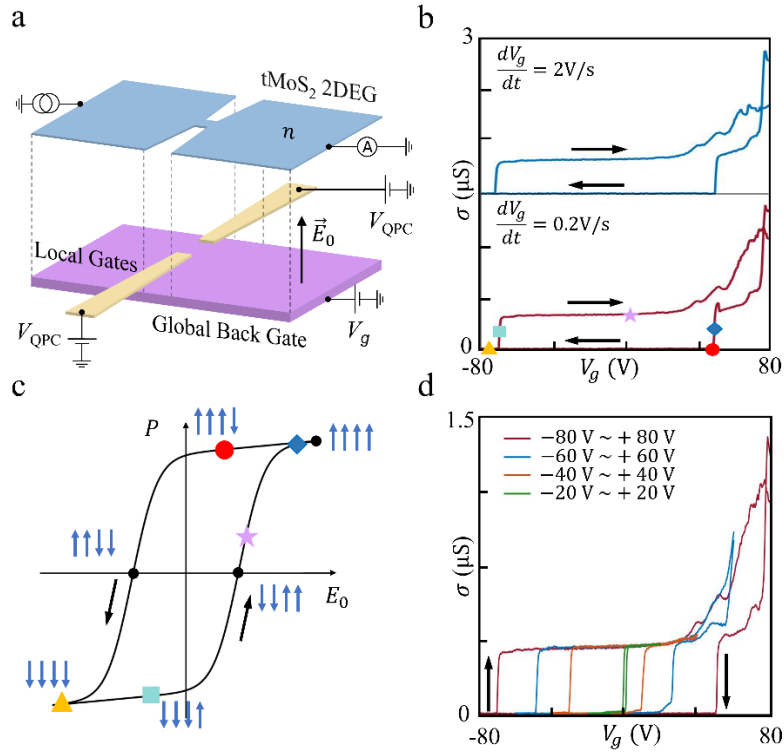

**Figure 2 | Electrical Hysteresis and Ferroelectric Polarization Dynamics in Twisted MoS$_2$ Quantum Point Contact.** (a) Schematic of local measurement scheme. The device consists of two local gates ($V_{QPC}$) and a global back gate ($V_g$) controlling the out-of-plane electric field $E_0$ and carrier density $n$ in 2DEG simultaneously. The tMoS$_2$ directly on top of the local gates are fully depleted by $V_{QPC}$, with the electrostatic-defined channel width on the same order as a single/few moiré domains. (b) Four-probe conductance measurements as a function of global back gate voltage $V_g$ at different ramp rates (blue line: fast; red line: slow), exhibiting pronounced hysteresis of the conductance plateau, independent of the scan rate. (c) Schematic of electric dipole polarization $P$ as a function of applied external field $E_0$. Blue arrows denote polarization directions. (d) Measured four-probe conductance as a function of global back gate voltage for forward and backward sweeps with the local gate fixed at $V_{QPC}$ = -7 V with different scan ranges.

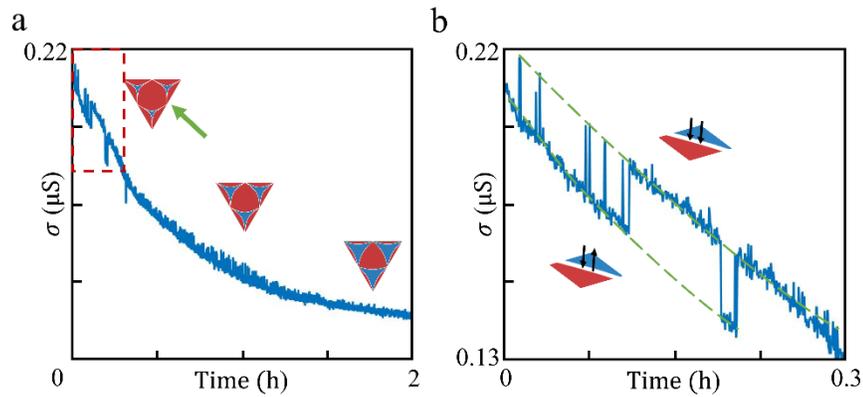

**Figure 3 | Quantum Point Contact as Single Atomic Dipole Sensor.** (a) Time evolution of the QPC conductance at the sharp slope between the first conductance plateau and pinch-off. Overall exponential decay of QPC conductance arises from evolution of adjacent ferroelectric moiré domain of opposite downward (blue) and upward (red) polarization, over a time scale of 2 hrs. (b) Finer scan over 0.3 hrs (marked by red dashed box in (a)). Discrete conductance spikes and steps are observed on top of the overall exponential behavior, with same quantized signal well above the measurement noise level, with green dashed lines marking the bottom and top of the spikes/steps. These conductance spikes/steps arise from random flipping of a single atomic dipole at a moiré domain boundary near the QPC. Insets: Schematics of atomic dipoles (black arrows) at the bottom and top of the spikes/steps.

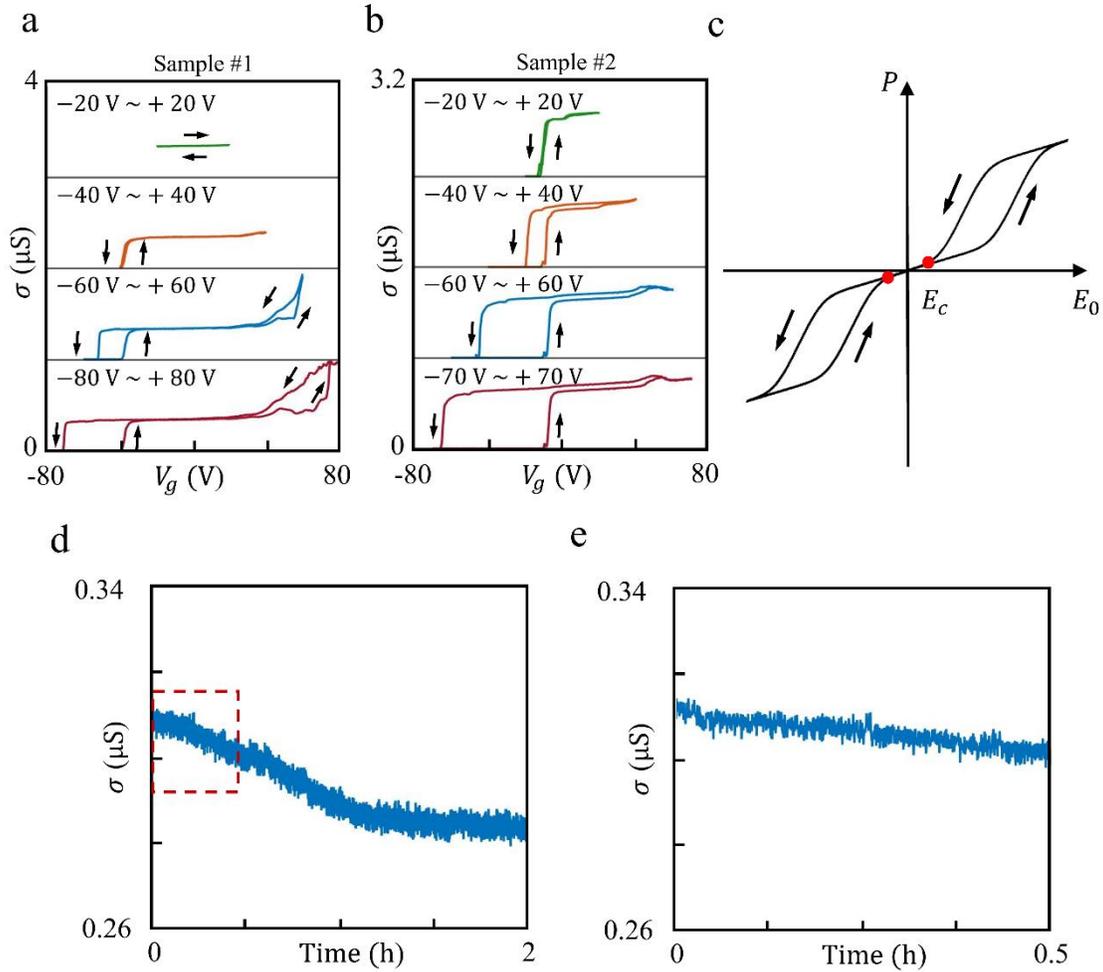

**Figure 4 | Anti-ferroelectric Behavior in Global Transport Measurements.** (a-b) Measured four-probe conductance via global measurements of (a) sample #1 and (b) sample #2, without the gate-defined QPCs, at different scan ranges. In contrast to the local measurement via gate-defined QPC, the global transport displays much smaller hysteresis loops located besides (instead of spanning over) $V_g = 0$, demonstrating anti-ferroelectric behaviors. (c) Schematic of anti-ferroelectric polarization as a function of the external electric field $E_0$, with critical field $E_c$ (for hysteresis to emerge) besides $V_g = 0$. (d) Time evolution of conductance via global transport measurements shows a similar domain evolution timescale, but absence of discrete spikes or steps in the (e) Zoom-in scan, without the single-dipole sensitivity from a QPC.

# Ferroelectric Quantum Point Contact in Twisted Transition Metal Dichalcogenides

Wei Ren[1], Shiyu Guo[1], Daochen Long[1], Noah Friedman[1], Jingyuan Xian[1], David Li[1], Jack Tavakley[1], Jeongsoo Park[1], Kenji Watanabe[2], Takashi Taniguchi[3], Ke Wang[1*]

[1]School of Physics and Astronomy, University of Minnesota, Minneapolis, MN, USA
[2]Research Center for Electronic and Optical Materials, National Institute for Materials Science, Tsukuba, Ibaraki, Japan
[3]Research Center for Materials Nanoarchitectonics, National Institute for Materials Science, Tsukuba, Ibaraki, Japan

## S1. Sample Preparation and Device Fabrication

The $tMoS_2$ stack is made following the 'tear-and-twist' method via the standard dry transfer technique [1] in a high-purity argon (99.9995%) environment to prevent the oxidization of $MoS_2$. Figure 1a is the schematic picture of the full heterostructure. With the help of a poly (bisphenol A carbonate) (PC) and polydimethylsiloxane (PDMS) stamp, a top hBN flake and a pair of few-layer graphite (FLG) are subsequently picked up. The two FLG flakes, pre-doped by a polypropylene carbonate (PPC) film, are used to improve the FLG contacts to $MoS_2$ [2]. Then two pieces of few-layer (up to five layers) $MoS_2$ flakes (figure 1a) are consecutively picked up with a relative twist angle targeting at < 1º. The stack is then encapsulated by a bottom hBN (55-75 nm thick) and transferred onto a set of pre-deposited atomically clean local gates (Cr/Pd-Au alloy, 1 nm/7 nm) on a $SiO_2$ (285 nm)/Si substrate. The doped Si substrate serves as a global back gate to tune the charge carrier density in the bulk $MoS_2$ via capacitive coupling. Two of the local bottom gates with a separation of 200 - 400 nm are utilized to form a QPC channel. Near the pinch-off point of the QPC, the width of the electrostatic defined conducting channel can be reduced to the order of a few tens of nanometers, which is comparable with the electron wavelength at typical carrier density. For small twist angles (< 1º), the moiré lattice constant is estimated to be a few tens of nm, thus the QPC region can hold approximately one or a few moiré domains. Electrical contacts to gates and ohmic contacts to 1D boundaries of FLG are made by electron-beam lithography, dry-etching and subsequent metal deposition (Cr/Pd/Au, 1 nm/5 nm/120 nm) [3]. A final round of dry-etching with an electron-beam lithography etch mask is performed to define the lateral geometry of the devices.

All measurement data is collected by a standard SR830 lock-in amplifier with an alternating voltage (AC) excitation of 1 mV at 17.7778 Hz in a Montana cryostat at a base temperature of 3.4 K unless otherwise specified. A Digital-to-Analog Converter (DAC) and/or Yokogawa DC voltage sources are used to apply voltages on the local metal gates, and a Keithley power supply is used to apply DC voltage on the global silicon back gate.

Most of the data presented in the main manuscript (figures 1,2,3, and 4a) is taken from tMoS$_2$ sample #1 with a 55 nm-thick bottom hBN. Similar results are observed in device regions D2 and D3 of tMoS$_2$ sample #2, following the same sample preparation and fabrication process as above but with a 75 nm-thick bottom hBN. Figure 4b is taken from device region D3 of the tMoS$_2$ sample #2. As a control, we also fabricate and measure a non-twisted few-layer MoS$_2$ QPC device with a 57 nm-thick bottom hBN.

## S2. Hysteresis Dependence on Scan Rate

In both tMoS$_2$ samples #1 and #2, no apparent scan rate dependence is observed across at least three orders of magnitude (from 0.02 V/s to 2 V/s) in the local measurement regime (figures 2b, S1a, S1b). This rules out the origin of hysteresis as the charging effects, either originating from the samples or the capacitive components in the measurement circuits. As a comparison, the conductance hysteresis in the non-twist MoS$_2$ device measured in the local measurement scheme yields significant scan rate dependence (figure S1c), which is typically attributed to the charge trap or disorder effect [4–6].

Additionally, there is no apparent dependence of hysteresis on scan rate in the global measurement scheme for both twisted MoS$_2$ sample #1 and sample #2 (figures S1d, e). No hysteresis is observed in the global measurement scheme for the non-twisted MoS$_2$ sample (figure S1f), indicating that typical hBN-encapsulated, microscale MoS$_2$ devices are sufficiently clean with respect to charge traps and disorder. Nonetheless, subtle local charge traps or disorder that can be revealed through the local measurement scheme (figure S1c) may persist beneath the globally uniform response.

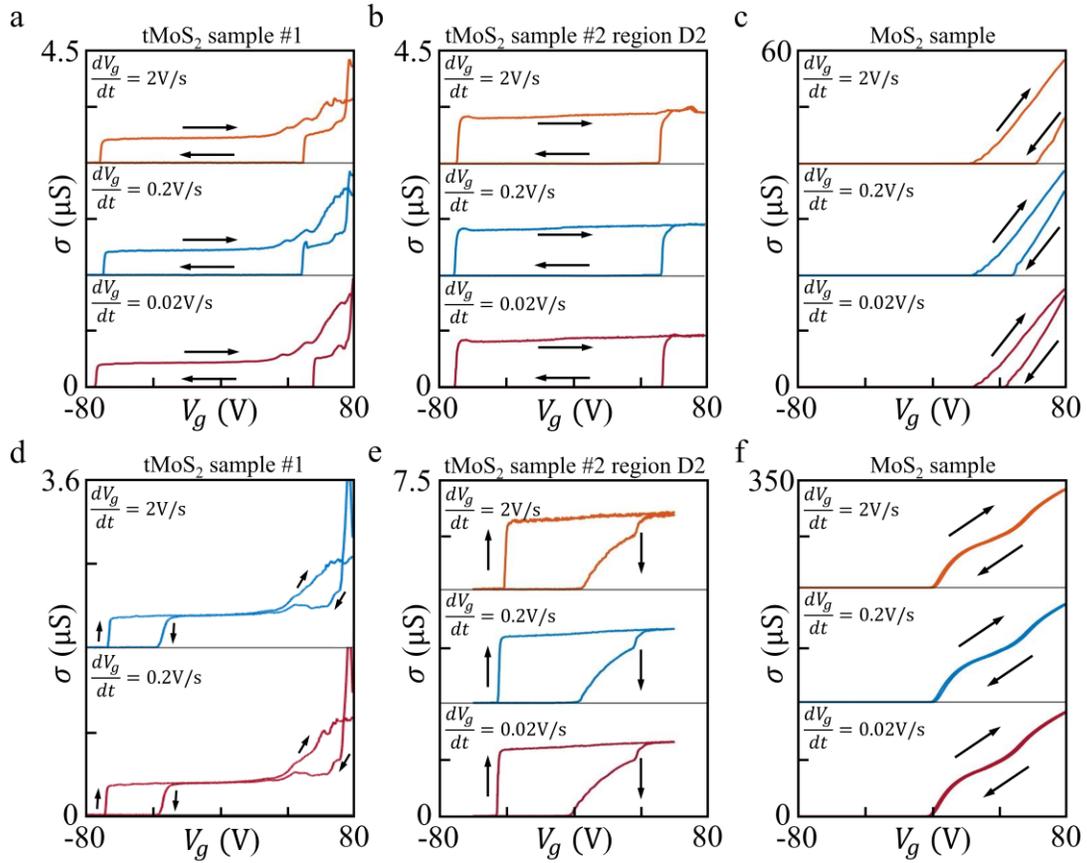

**Figure S1. Scan Rate Dependence of Hysteresis Loop.** Measured conductance hysteresis at three different scan rates in the local measurement scheme in (a) device region D1 in tMoS$_2$ sample #1 (b) device region D2 of tMoS$_2$ sample #2, and (c) a non-twist MoS$_2$ sample. (d-f) Conductance hysteresis probed by the global measurements in (d) tMoS2 sample #1 (e) device region D2 of tMoS$_2$ sample #2 (f) a non-twist MoS$_2$ sample.

## S3. Capacitive Coupling Ratio Calculation

The tMoS$_2$ devices are effectively bottom-gated field effect transistors, where the carrier density $n$ in the bulk 2DEG is determined by

$$n = \frac{c_g V_g}{e}, \tag{1}$$

where $c_g$ is the capacitance per unit area between the silicon back gate and tMoS$_2$ 2DEG, given by

$$c_g = \frac{c_{hBN} c_{SiO_2}}{c_{hBN} + c_{SiO_2}}. \tag{2}$$

Here, $c_{SiO_2}$ and $c_{hBN}$ are the capacitance per unit area corresponding to the bottom hBN layer and SiO$_2$ substrate underneath tMoS$_2$, respectively, which are calculated given the thickness and permittivity of both materials.

From Gauss's law, the electric field in hBN, $E_0$, is given by

$$E_0 = \frac{ne}{\varepsilon_{hBN}\varepsilon_0} = \frac{c_g}{\varepsilon_{hBN}\varepsilon_0}V_g, \quad (3)$$

where $\varepsilon_{hBN}$ is the out-of-plane relative permittivity of hBN. The charge carrier density is therefore related to $E_0$ through relation

$$n = \frac{\varepsilon_{hBN}\varepsilon_0}{e}E_0. \quad (4)$$

Considering the contribution from the molecular field, $E'$, the charge carrier density can be related to the net electric field, $E_{\text{net}}$, by

$$n = CE_{\text{net}} \equiv C(E_0 - E'), \quad (5)$$

where $C$ is a prefactor that determined by the dielectric environment of 2DEG.

For the 2DEG region directly above the local gates, the carrier density is given by

$$n = \frac{c_{hBN}V_{QPC}}{e}, \quad (6)$$

as the bottom hBN is the only dielectric layer between the local gates and tMoS$_2$ 2DEG.

In order to measure the sample as a globally uniform 2DEG (same change in $n$ across the entire sample), the ratio of the global back gate voltage, $\Delta V_g$, and the local bottom gate voltage, $\Delta V_{QPC}$, should satisfy:

$$\Delta n = \frac{c_g \Delta V_g}{e} = \frac{c_{hBN}\Delta V_{QPC}}{e}, \quad (7)$$

so that

$$\Delta V_g / \Delta V_{QPC} = c_{hBN}/c_g.$$

## S4. Absence of Hysteresis in Local Gate Scans

Figure S2a is the four-probe conductance as a function of local gate voltage $V_{\text{QPC}}$ at a fixed back gate voltage $V_g$ = +45 V, measured in both forward and backward scan directions. The hysteresis is nearly vanishing, which is explained as the following. Near the pinch-off point, the tMoS$_2$ directly on top of the local gates do not participate in the transport. The QPC channel is fine-tuned by the fringing fields from the local gates from the side (figure S2b). The fringing fields primarily extend laterally and are therefore less effective at changing the total dipole moment of the ferroelectric domain within the QPC channel, in contrast to the back gate voltage, which directly influences the domain from beneath. As a result, the hysteresis loop is significantly more suppressed compared to that observed under global back gate tuning.

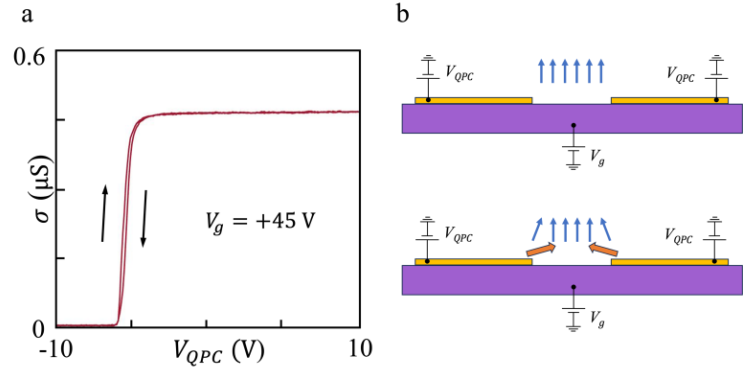

**Figure S2. Absence of Hysteresis in Local Gate Scans.** (a) Measured conductance as a function of $V_{QPC}$ at a fixed back gate voltage $V_g$ = +45 V. (b) Schematic of the effects of fringing fields (orange arrows) from the local gates. Only the dipole moments (blue arrows) near the two edges of the QPC channel are partially affected by the local gates, which are canted towards opposite direction, resulting in a minimal change in the total dipole moments in the QPC channel.

## S5. Long Conductance Plateau in Global Measurements

Although shorter than the unusually long conductance plateau measured in the local measurement scheme (figure 2b), a long conductance plateau is also present in the global measurement scheme figures 4a, b. This may be attributed to the finite slope in the antiferroelectric polarization loop (figure 4c), $P(E_0)$ near $E_0 = 0$, or below the critical electric field. Although there is no hysteresis behavior, the bottom gates (both Si back gate and local gates) are ineffective in tuning the carrier density in this regime due to the presence of finite polarization or molecular fields. It may also arise from the existence of large ferroelectric domains in the measured region due to twist angle inhomogeneity [7,8], which contribute significantly to the observed long conductance plateau. Further experimental evidence such as TEM imaging is needed to confirm the nature of long conductance plateau.

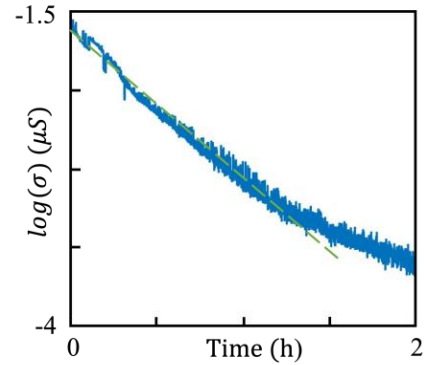

**Figure S3. Time Evolution of QPC Conductance on a Semi-log Scale.** Time-dependence profile from figure 3a, replotted on a semi-log scale, exhibits a slight upward deviation from a straight line (green dashed line).

## S6. Time Evolution of QPC Conductance on a Semi-log Scale

The same time-dependence as in figure 3a is replotted on a semi-log scale here (figure S3). The slight slow-down compared with the exponential trend suggests that the electric dipoles in tMoS$_2$ are weakly interacting or largely independent.

## S7. Results from Multiple Devices

Figure S4c and figure 4b show the conductance and hysteresis loop measured via the global measurements in device regions D2 and D3 of sample #2, respectively. Note that figure S4c is taken with an AC voltage excitation of 0.1 mV, all other data is taken under 1 mV AC voltage excitation. The conductance, rather than forming a long, perfectly flat plateau, exhibits a slight uptick in the global measurements, suggesting that molecular field effects are slightly suppressed by averaging over alternating ferroelectric domains. The non-twist MoS$_2$ sample does not have any long conductance plateau, confirming that the usually long plateau originates from the moiré ferroelectrics (figure S1f). The local measurement results are summarized in figures S4a, b, with additional tMoS$_2$ devices exhibiting performance that qualitatively agrees with sample #1 in the main text.

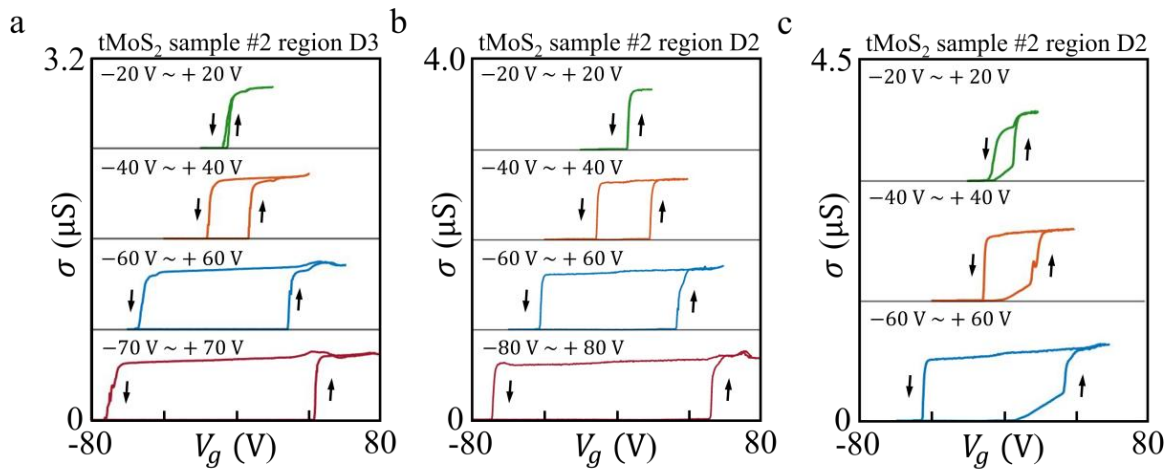

**Figure S4. Hysteresis Measurements in Additional Devices.** (a-b) Measured four-probe conductance via local measurements of (a) device region D3 and (c) device region D2 of sample #2 at different scan ranges. (c) Measured four-probe conductance via global measurements of device region D2 of sample #2, without the gate-defined QPCs, at different scan ranges.

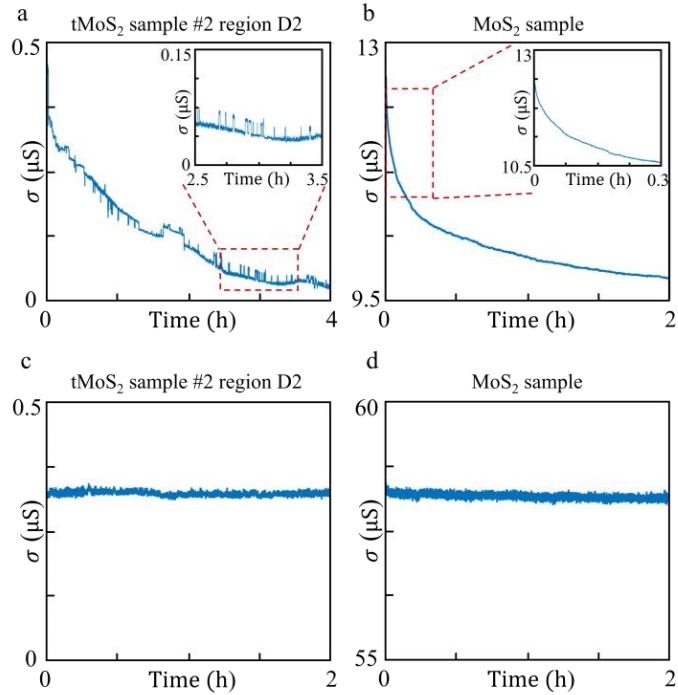

**Figure S5. Time Evolution of Ferroelectric Domains in Additional Devices.** (a-b) Time evolution of measured conductance when parking at half-way towards the conductance plateau probed via local measurements in (a) device region D2 of tMoS$_2$ sample #2 and (b) non-twist MoS$_2$ sample. Insets are zoom-in scans. Only tMoS$_2$ samples have spikes and steps on top of the exponential decay profile, indicating the origin of the spikes being single dipole flipping events. (c-d) Time evolution of conductance probed via global measurements for (c) device region D2 of tMoS$_2$ sample #2 and (d) the non-twisted MoS$_2$ sample. These measurements reveal negligible temporal variation over timescales of several hours. Notably, the presence of conductance spikes is significantly suppressed.

Figures S5a presents time-domain measurements taken from device region D2 of tMoS$_2$ sample #2. The local measurement exhibits a characteristic near-exponential decay, consistent with the relaxation dynamics of weakly interacting dipoles approaching equilibrium, which is indicative of single-domain behavior. The inserted figure is the 1-hr zoom-in on long-time scale measurements. The telegraph like switching behavior between two distinct floor values was detected, consistent with the single dipole flipping at the domain boundary near the QPC. These switching events demonstrate sensitivity to single-dipole dynamics, which match the main device (sample #1) results as shown in figure 3. Global measurements exhibit stable conductance (figure S5c) with negligible temporal evolution. The disappearance of telegraph-like switching behavior in global measurements further suggests that the conductance is not dominated by the local dingle dipole dynamics but rather indicates the averaged behavior, insensitive to the evolution of local dipoles.

Figure S5d is the time evolution measured in the few-layer untwisted $MoS_2$ device. Local measurement shows exponential conductance decay without the telegraph-like switching behavior, indicating that the observed switching events in twisted $MoS_2$ in sample #1 and sample #2 arises from single-dipole flipping unique to the moiré ferroelectric system.